\documentclass[journal]{IEEEtran}
\usepackage{amsmath,amssymb,amsfonts}
\usepackage{array}
\usepackage{siunitx}
\usepackage[caption=false,font=normalsize,labelfont=sf,textfont=sf]{subfig}
\usepackage{textcomp}
\usepackage{booktabs}
\usepackage{graphicx}
\usepackage{xcolor}
\usepackage{url}
\usepackage{eurosym}
\usepackage{mathtools}
\usepackage{orcidlink}
\usepackage{hyperref}

\usepackage[noadjust]{cite}

\graphicspath{{../Figures/}{Figures/}}

\begin{document}

\title{A Benchmarking Case Study for Local Flexibility Markets: Network, Scenarios, and Open Inputs}

\author{
Savvas~Panagi\,\orcidlink{0009-0008-6057-3392},~\textit{Student Member, IEEE},
Chrysovalantis~Spanias\,\orcidlink{0000-0003-3046-3287},~\textit{Member, IEEE},
and Petros~Aristidou\,\orcidlink{0000-0003-4429-0225},~\textit{Senior Member, IEEE}
\thanks{S. Panagi and C. Spanias are with the Distribution System Operator, Electricity Authority of Cyprus, Nicosia, Cyprus.}
\thanks{S. Panagi and P. Aristidou are with the Department of Electrical \& Computer Engineering \& Informatics, Cyprus University of Technology, Limassol, Cyprus.}
\thanks{Network data, load profiles, flexibility offers, and parameter files are available in an open-access Zenodo repository~\cite{panagi2026zenodo}.}
}

\maketitle

\begin{abstract}
External inputs to local flexibility markets, network models, demand scenarios, flexibility offers, and asset thermal, economic parameters do not change the clearing methodology, yet they are routinely defined under heterogeneous assumptions, which hinders reproducible benchmarking. This paper provides a complete, methodology-agnostic input set built around a modified CIGRE MV network with prescribed base-load and stress-load conditions, synthetic flexibility offers, wholesale prices, and transformer and cable parameters. Baseline AC operating points (voltages, loadings, and aging) are reported for both days so that alternative flexibility-request and clearing methods can be applied to the same inputs and compared fairly. The associated data are released openly on Zenodo.
\end{abstract}


\section{Introduction}
\label{sec:supp_intro}

Local flexibility markets (LFMs) separate, to varying degrees, the market-clearing mechanism from the network-specific inputs that drive flexibility valuation. The clearing formulation itself, objective structure, offer acceptance rules, and privacy constraints, can be studied independently of any particular feeder. In contrast, the \emph{external inputs} that feed the clearing problem, including the network model, baseline and stressed demand profiles, flexibility offers, wholesale prices, and the thermal and economic parameters of transformers and cables, do not alter the methodology but strongly affect numerical outcomes. Across the literature these inputs are typically assembled under heterogeneous and often incompletely documented assumptions, which limits reproducibility and makes fair benchmarking between flexibility-request methods difficult.

This paper addresses that gap by defining a complete, openly specified input set for LFM benchmarking. We adopt a modified CIGRE European MV network, prescribe base-load and stress-load (EV) operating conditions, publish synthetic but fully stated flexibility offers, and fix ambient temperature together with transformer and cable thermal--economic parameters. All network data, time series, and parameter files are released in an open-access Zenodo repository~\cite{panagi2026zenodo}. Baseline AC power-flow results (voltages, loadings, and aging) are reported without assuming any particular clearing method, so that independent groups can apply their own flexibility-request or market-clearing formulations to the same inputs and compare outcomes on equal footing.

The remainder of the paper is organized as follows. Section~\ref{sec:supp_case_study} specifies the network, scenarios, and numerical inputs.
Section~\ref{sec:supp_baseline_results} reports the baseline operating points for the base and stress days.
Section~\ref{sec:supp_data} describes the Zenodo release.
Section~\ref{sec:supp_conclusion} concludes.

\section{Case Study: Network, Scenarios, and Inputs}
\label{sec:supp_case_study}

This section defines the network, operating scenarios, and numerical inputs used for benchmarking. The same specification underlies the numerical experiments reported in~\cite{panagi2026dynamic}; the machine-readable files are released on Zenodo~\cite{panagi2026zenodo}.

\subsection{Network Model and Modifications}
\label{sec:supp_network}

\begin{figure}[b!]
    \centering
    \includegraphics[width=0.95\linewidth]{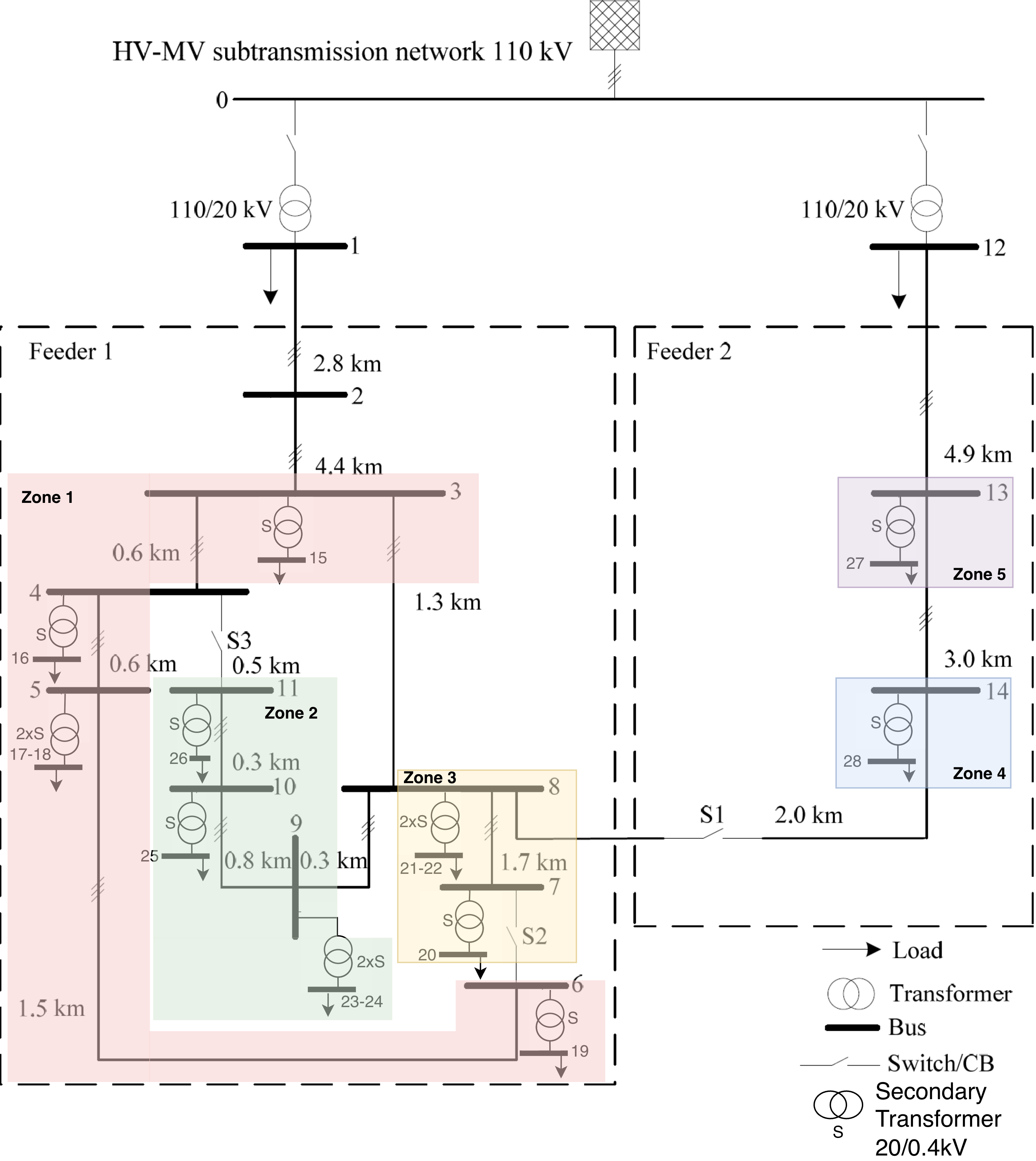}
    \caption{CIGRE single-line diagram of the studied 20-kV distribution system, including the primary and secondary substations.}
    \label{fig:supp_cigre_network}
\end{figure}

The case study is built on the standard CIGRE European MV benchmark network as depicted in Fig. \ref{fig:supp_cigre_network}. The original model represents a two-feeder meshed $110/20$~kV system with aggregated MV loads, although the network is operated radially, with the tie switches open. For the present study, the network is modified such that each controllable MV load point is represented by a dedicated $20/0.4$~kV secondary substation. Two MV buses, buses $1$ and $12$, are excluded from this secondary-substation expansion. These buses represent large, non-controllable MV demand points directly connected to the two primary feeders and remain fixed throughout the study.

Following these modifications, the network contains two primary transformers and 14 secondary substations, as shown in Fig.~\ref{fig:supp_cigre_network}. All simulations are performed over a $24$-hour horizon with hourly time steps. Because some flexibility-request methodologies partition buses into zones in order to enlarge the set of eligible providers and thereby increase market liquidity, e.g.,~\cite{prat-network-aware}, the benchmarking setup also defines five indicative zones as illustrated in Fig.~\ref{fig:supp_cigre_network}.

\subsection{Operating Scenarios}
\label{sec:supp_scenarios}

\begin{figure}[b!]
    \centering
    \includegraphics[width=1\linewidth]{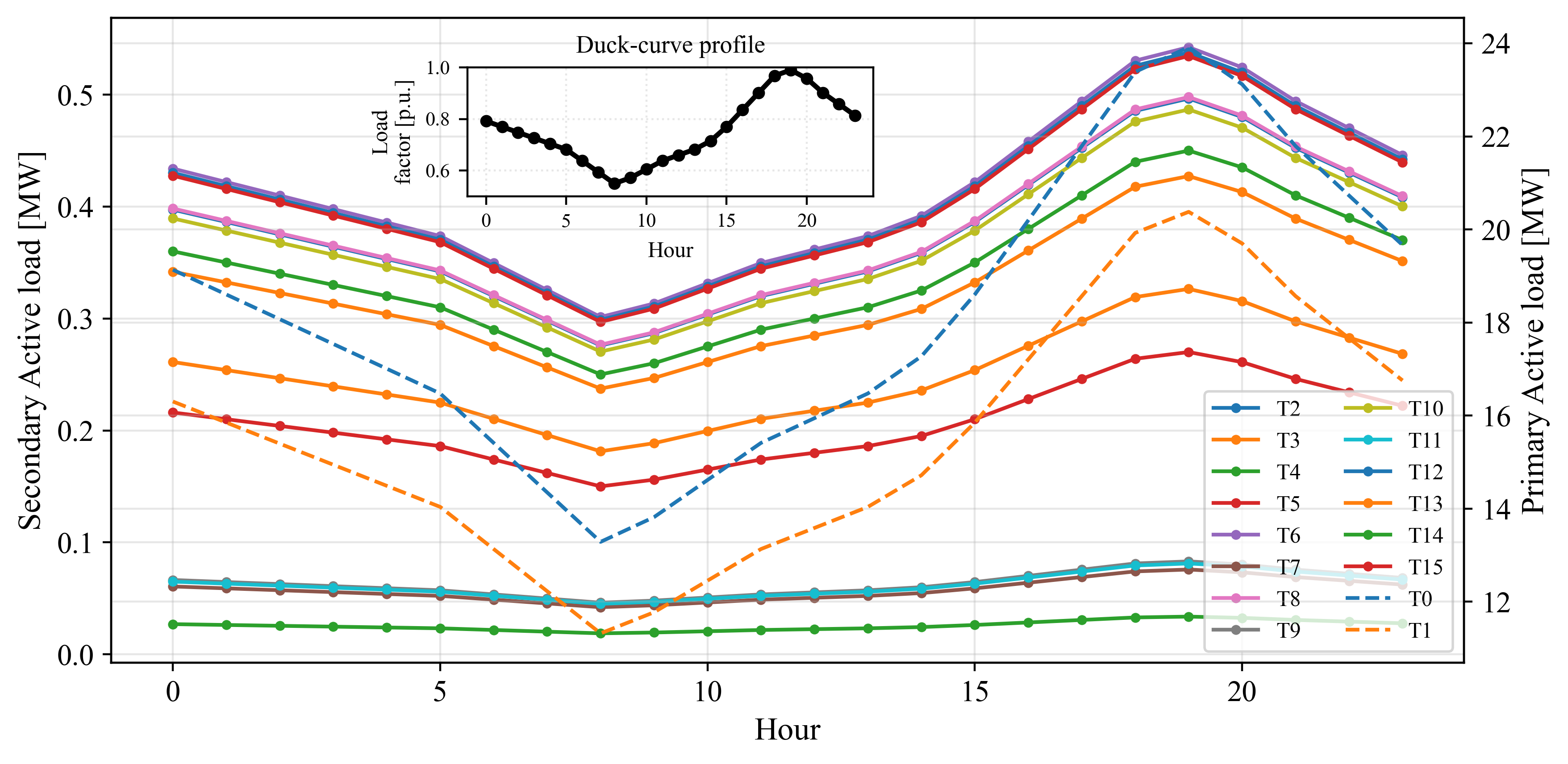}
    \caption{Base-case demand: common duck-curve load factor and the corresponding active-power profiles at the primary and secondary transformer feeders.}
    \label{fig:supp_base_load_profiles}
\end{figure}

\begin{figure}
    \centering
    \includegraphics[width=1\linewidth]{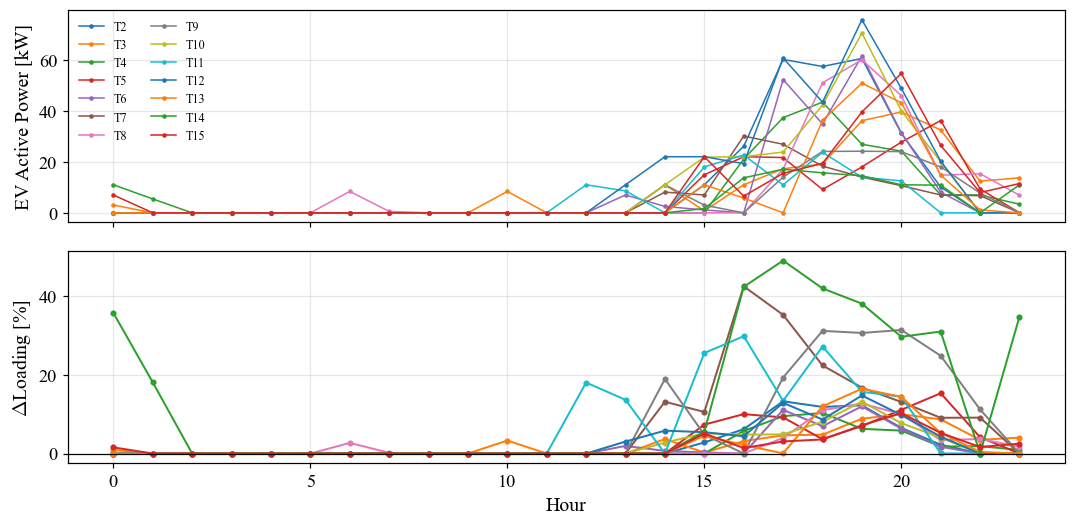}
    \caption{Stress case: uncontrolled EV active power at each secondary substation (top) and the resulting relative secondary-transformer loading increase $\Delta\mathrm{Loading}$ relative to the base case (bottom).}
    \label{fig:supp_stress_load_profiles}
\end{figure}

Two operating scenarios are considered. The \textit{base case} represents a typical suburban load day. A common normalized \textit{duck-curve} active-load profile (in p.u.) is applied identically to every load connected to the network; the resulting primary- and secondary-transformer feeder active-power trajectories are shown in Fig.~\ref{fig:supp_base_load_profiles}. Because all loads share the same shape, the feeder time series differ only through their rated demand levels. 
The \textit{stress case} retains this duck-curve base demand but superimposes uncontrolled residential EV charging at each secondary substation. The EV load follows the statistical assumptions used in~\cite{11305421}. Each secondary substation serves $30$--$50$ customers based on their nominal apparent power, of which $30\%$ own an EV. Connection type is assigned probabilistically, with $80\%$ to three-phase supply and $20\%$ to single-phase supply. The corresponding home chargers are rated at $11$~kW (three-phase) and $7$~kW (single-phase). Each EV charges at its rated charger power until the daily trip energy is restored. Fig.~\ref{fig:supp_stress_load_profiles} shows the resulting EV active-power consumption at each secondary and the relative increase in secondary-transformer loading, which is concentrated in the evening charging window.

\subsection{Input Parameters}
\label{sec:supp_inputs}

\subsubsection{Transformer aging parameters}
Transformer aging follows the IEEE~C57.91 Arrhenius model~\cite{6166928}. The rated temperature rises are $\Delta\theta^{TO}_{R} = \SI{55}{\celsius}$ and $\Delta\theta^H_{R} = \SI{25}{\celsius}$ \cite{andrianesis2021impact}. Oil and winding exponents are $n=m=0.8$ \cite{6166928}; the top-oil and hotspot time constants are $3$~h and $4$~min \cite{andrianesis2021impact}, respectively. The normal life expectancy is taken as \SI{180000}{h} \cite{6166928}, and the replacement cost is assumed proportional to rated power at \SI{100000}{EUR/MVA} \cite{1668461}.

\subsubsection{Cable aging parameters}
Cable aging uses an Arrhenius conductor-temperature model with parameters from~\cite{andrianesis2021impact,energy2015climate}: rated surface and core temperature rises $\Delta\theta_{CS,R} = \SI{38}{\celsius}$ and $\Delta\theta_{C,R} = \SI{40}{\celsius}$, time constants $\tau_{CS} = \SI{2}{h}$ and $\tau_C = \SI{4}{min}$, Arrhenius coefficient $B = 15{,}000$, normal cable life \SI{438000}{h}, and replacement cost \SI{80000}{EUR/km} multiplied by the installed cable length.

\subsubsection{Market and network assumptions}
\begin{figure}[b!]
    \centering
    \includegraphics[width=1\linewidth]{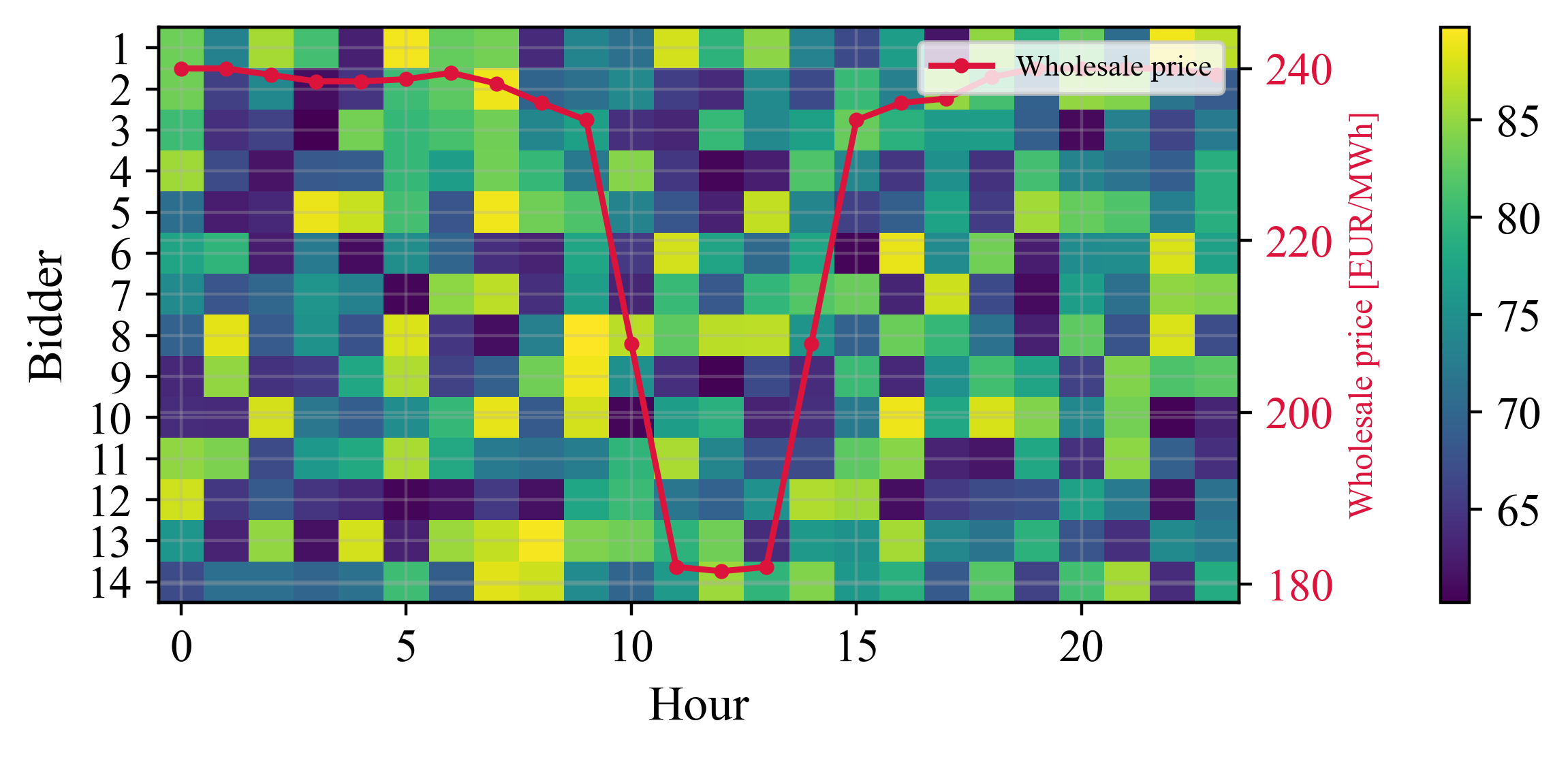}
    \caption{Wholesale day-ahead price and FSP hourly bid prices by bidder.}
    \label{fig:supp_fsp_bids}
\end{figure}

The remaining inputs are network limits and exogenous price/offer data. Voltage limits are $\underline{V} = 0.90$~p.u. and $\overline{V} = 1.10$~p.u. The baseline AC operating point assumes a constant load power factor of $0.95$ (lagging). Wholesale energy prices follow the Cyprus day-ahead clearing price on 18~July~2026~\cite{TSOC_DAM_Prices}. Fourteen local FSPs (one per secondary substation) submit synthetic hourly offers with capacities of $120$~kW and bid prices drawn uniformly from $60$--$90$~\euro/MWh; the wholesale series and the resulting bid-price heatmap are shown in Fig.~\ref{fig:supp_fsp_bids}. The inputs are summarized in Table \ref{tab:supp_param_summary}.

\begin{table}
\centering
\caption{Summary of key benchmarking parameters.}
\label{tab:supp_param_summary}
\begin{tabular}{@{}ll@{}}
\toprule
Quantity & Value \\
\midrule
Horizon / resolution & $24$~h / $1$~h \\
Primary transformers & $2$ \\
Secondary substations / FSPs & $14$ / $14$ \\
FSP capacity (each) & $120$~kW \\
FSP bid-price range & $60$--$90$~\euro/MWh \\
Load power factor & $0.95$ (lagging) \\
Voltage limits & $0.90$--$1.10$~p.u. \\
Transformer replacement cost & \SI{100000}{EUR/MVA} \\
Cable replacement cost & \SI{80000}{EUR/km} \\
\bottomrule
\end{tabular}
\end{table}

\section{Baseline Operating-Point Results}
\label{sec:supp_baseline_results}

Given the network, demand scenarios, and asset parameters of Section~\ref{sec:supp_case_study}, a full AC power flow is solved independently for each hour of the base and stress days. No flexibility activation is considered in this section: the purpose is to document the resulting operating points so that any flexibility-request or clearing methodology can be applied to the same inputs and compared on equal footing.

\subsection{Voltage and thermal loading}
\label{sec:supp_v_loading}

\begin{figure}
    \centering
    \includegraphics[width=0.95\linewidth]{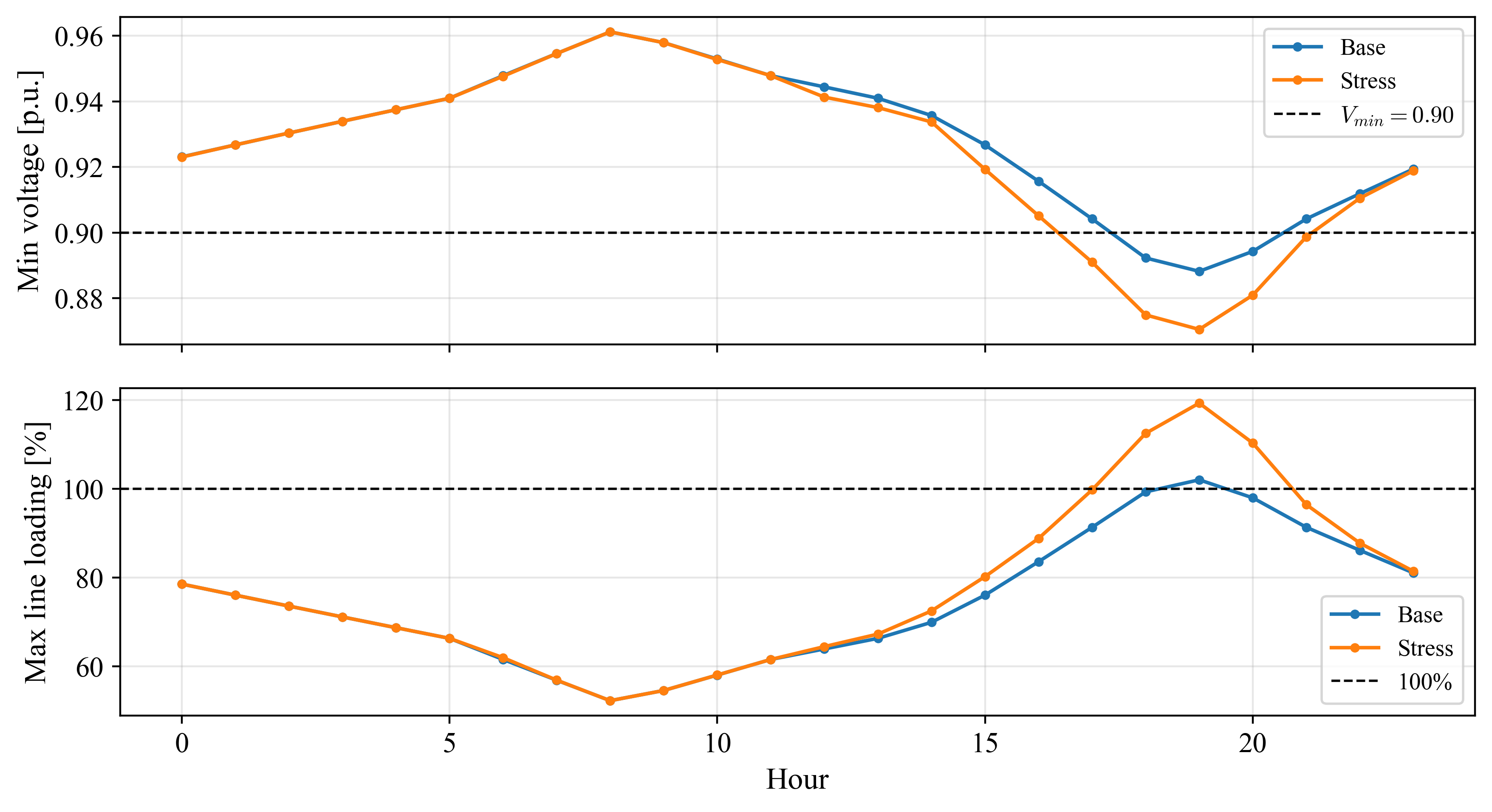}
    \caption{Baseline AC operating point: minimum feeder voltage (top) and maximum line loading (bottom) for the base and stress cases. Dashed lines mark $\underline{V}=0.90$~p.u.\ and $100\%$ loading.}
    \label{fig:supp_baseline_voltage_loading}
\end{figure}

Fig.~\ref{fig:supp_baseline_voltage_loading} compares the network-wide minimum bus voltage and the maximum line loading over the day. In the base case, the evening peak drives the minimum voltage slightly below $\underline{V}=0.90$~p.u.\ at hours~$18$--$20$ (lowest value $\approx 0.89$~p.u.\ at hour~$19$), while the maximum line loading exceeds $100\%$ only at hour~$19$ ($\approx 102\%$). Under the stress case, uncontrolled EV charging deepens and prolongs these violations: the minimum voltage falls to $\approx 0.87$~p.u.\ at hour~$19$ and remains below $0.90$~p.u.\ over hours~$17$--$21$, and the maximum line loading reaches $\approx 119\%$ at hour~$19$, with overload over hours~$18$--$20$.

\subsection{Transformer and cable aging at the baseline}
\label{sec:supp_aging_baseline}

\begin{figure}
    \centering
    \includegraphics[width=0.95\linewidth]{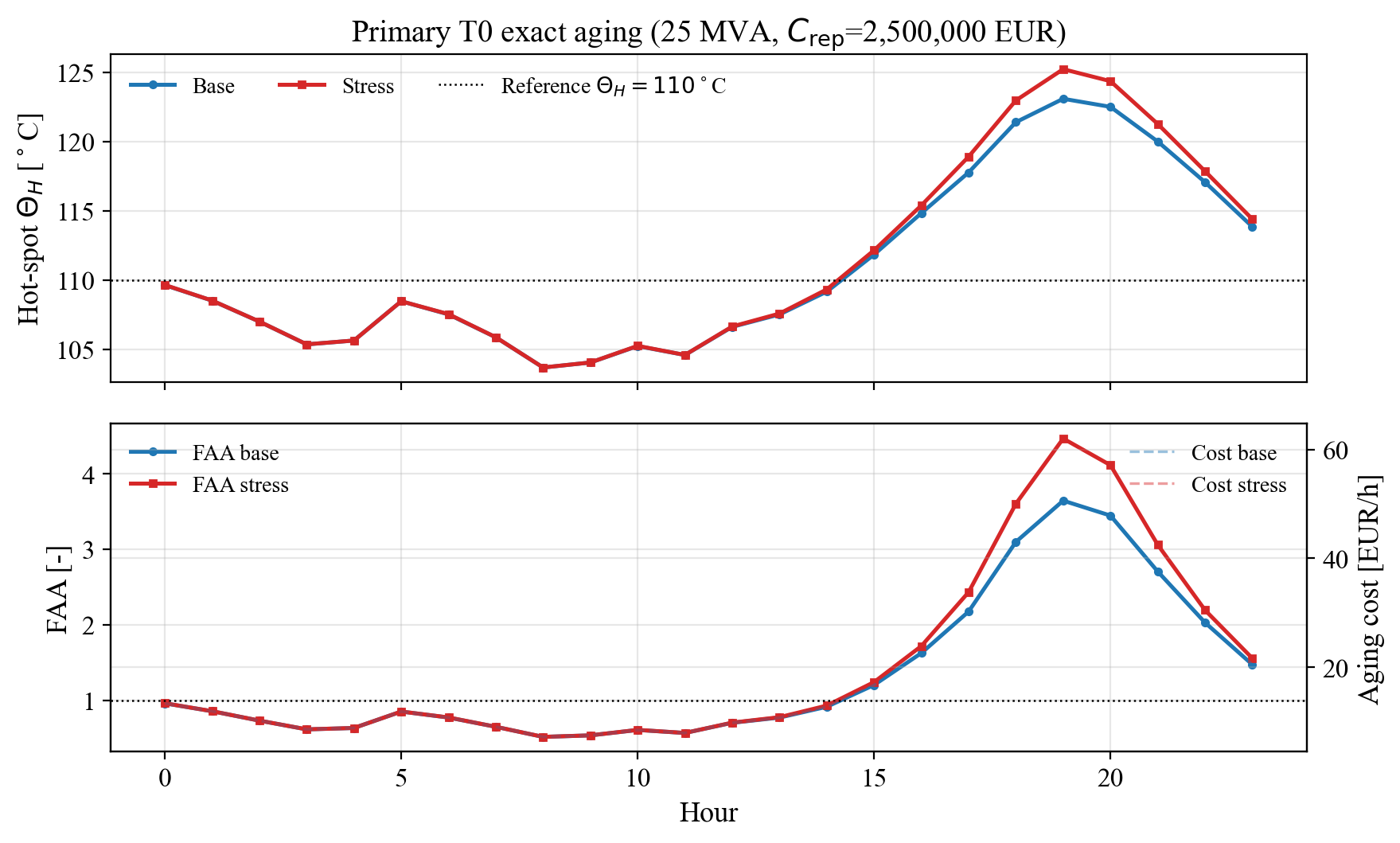}
    \caption{Exact IEEE~C57.91 aging of primary transformer~T0 under the base and stress baselines: hot-spot temperature (top) and FAA with hourly aging cost (bottom).}
    \label{fig:supp_transformer_aging}
\end{figure}

\begin{figure}
    \centering
    \includegraphics[width=0.95\linewidth]{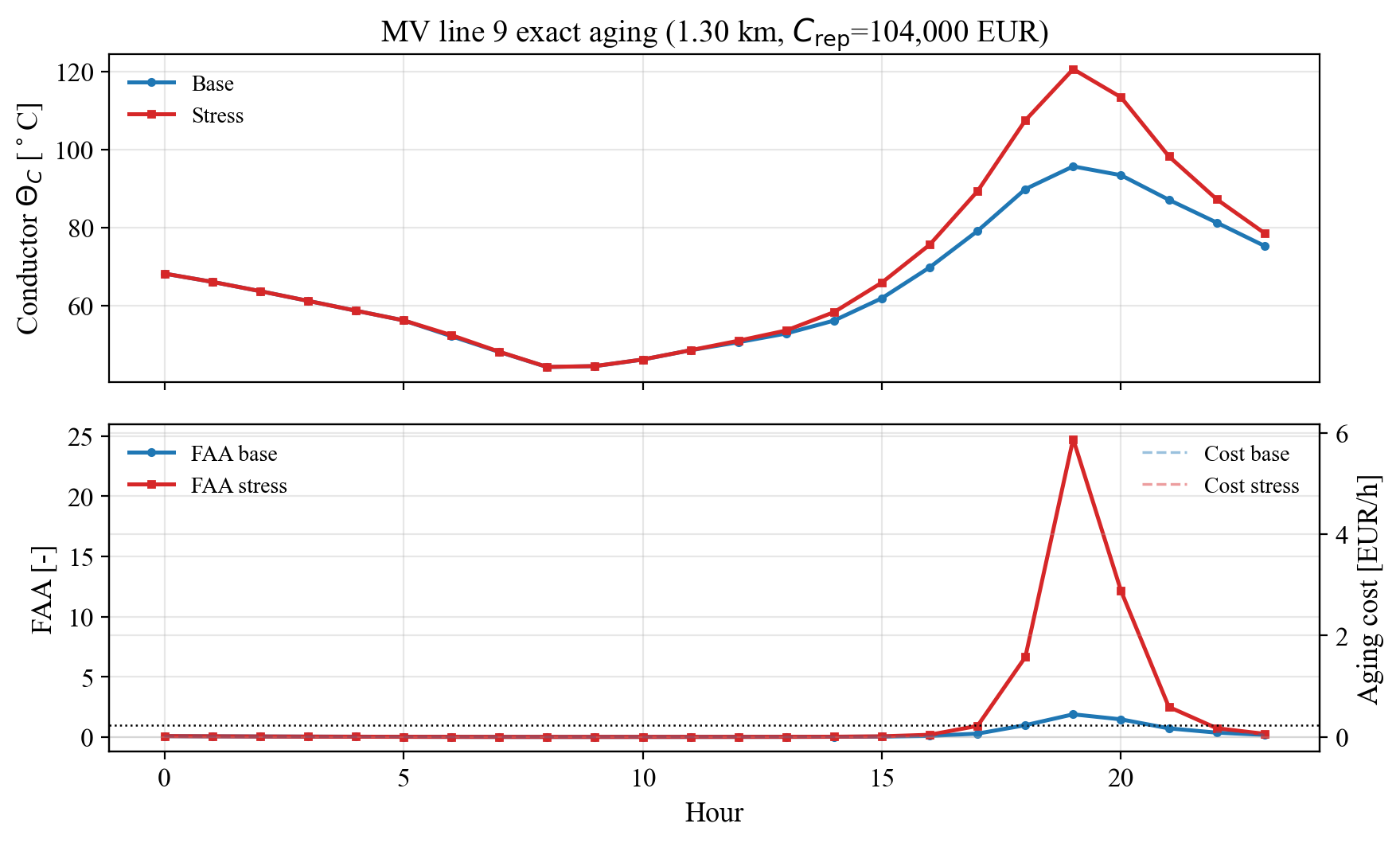}
    \caption{Exact transient Arrhenius aging of the most loaded MV cable (line~$9$) under the base and stress baselines: conductor temperature (top) and FAA with hourly aging cost (bottom).}
    \label{fig:supp_cable_aging}
\end{figure}

With the thermal--economic parameters of Section~\ref{sec:supp_case_study}, the exact nonlinear aging models are evaluated on the baseline loadings $K_{\mathrm{tra},t}$ and $K_{l,t}$ (no flexibility activation). Fig.~\ref{fig:supp_transformer_aging} shows primary transformer~T0 ($25$~MVA): the peak hot-spot temperature rises from $123.1\,^\circ\mathrm{C}$ (base) to $125.2\,^\circ\mathrm{C}$ (stress), the peak FAA from $3.64$ to $4.47$, and the daily aging cost from $446$ to $488$~\euro\ ($+9.4\%$). Fig.~\ref{fig:supp_cable_aging} reports the most loaded MV cable under stress (line~$9$, $1.3$~km): conductor temperature reaches $120.6\,^\circ\mathrm{C}$ versus $95.6\,^\circ\mathrm{C}$ in the base day, peak FAA jumps from $1.88$ to $24.8$, and the daily aging cost from $1.5$ to $11.5$~\euro. Aggregating over all MV lines, the daily cable aging cost increases from about $10$ to $49$~\euro. These trajectories provide a shared numerical reference for methods that monetize asset degradation on this benchmark.

\section{Data Availability}
\label{sec:supp_data}

The complete benchmarking package, including the modified CIGRE network in a pandapower format, base- and stress-load profiles, flexibility-offer sets, wholesale price series, and transformer/cable thermal and economic parameters, is available under open access on Zenodo~\cite{panagi2026zenodo}. The repository is intended to support reproducible comparison of LFM methods under a shared external-input specification.

\section{Conclusion}
\label{sec:supp_conclusion}

This paper provides an openly specified benchmarking package for local flexibility markets: a modified CIGRE MV network, base and stress demand scenarios, FSP offers, wholesale prices, and transformer/cable thermal--economic parameters, together with the associated baseline AC operating points (voltages, loadings, and aging). Because the package is methodology-agnostic, alternative flexibility-request and clearing formulations can be applied to the same inputs and compared fairly. All data are released on Zenodo~\cite{panagi2026zenodo} to facilitate reuse and extension.

\bibliographystyle{IEEEtran}
\bibliography{biblio_supp}

\end{document}